\documentclass[aps,prr,reprint,twocolumn,superscriptaddress,floatfix,nofootinbib]{revtex4-1}
\usepackage{amsmath,amssymb,color,comment}
\usepackage[makeroom]{cancel}
\usepackage[caption=false]{subfig}
\usepackage{mathrsfs}
\usepackage{graphicx}
\usepackage{subfig}
\usepackage[countmax]{subfloat}
\usepackage[english]{babel}
\usepackage[normalem]{ulem}
\definecolor{jadclr}{rgb}{0,0.5,0}
\definecolor{jadcolor}{rgb}{0.5,0,0}
\usepackage[bookmarks=true,colorlinks,linkcolor=jadclr,urlcolor=jadcolor,citecolor=jadcolor]{hyperref}
 \usepackage{dsfont}

\definecolor{mypink1}{rgb}{0.858, 0.188, 0.478}

\begin{document}

\title{Quasiparticle origin of dynamical quantum phase transitions}
\author{Jad C.~Halimeh}
\affiliation{Max Planck Institute for the Physics of Complex Systems, 01187 Dresden, Germany} 
\affiliation{Physics Department, Technical University of Munich, 85747 Garching, Germany}

\author{Maarten Van Damme}
\affiliation{Department of Physics and Astronomy, University of Ghent, Krijgslaan 281, 9000 Gent, Belgium}

\author{Valentin Zauner-Stauber}
\affiliation{Vienna Center for Quantum Technology, University of Vienna, Boltzmanngasse 5, 1090 Wien, Austria}

\author{Laurens Vanderstraeten}
\affiliation{Department of Physics and Astronomy, University of Ghent, Krijgslaan 281, 9000 Gent, Belgium}

\begin{abstract}
Considering nonintegrable quantum Ising chains with exponentially decaying interactions, we present matrix product state results that establish a connection between low-energy quasiparticle excitations and the kind of nonanalyticities in the Loschmidt return rate. When domain walls in the spectrum of the quench Hamiltonian are energetically favored to be bound rather than freely propagating, anomalous cusps appear in the return rate regardless of the initial state. In the nearest-neighbor limit, domain walls are always freely propagating, and anomalous cusps never appear. As a consequence, our work illustrates that models in the same equilibrium universality class can still exhibit fundamentally distinct out-of-equilibrium criticality. Our results are accessible to current ultracold-atom and ion-trap experiments.
\end{abstract}

\date{\today}
\maketitle

\section{Introduction}
It is no overstatement that critical phenomena are among the most intriguing and actively investigated subjects in physics, and have been extensively studied theoretically and experimentally for decades. Theoretical understanding has been established by the renormalization-group method, which relates criticality to scale invariance, universality, and a characteristic set of critical exponents \cite{Sachdev_book, Ma_book, Cardy_book}. A natural question, motivated by substantial technological advancement on the experimental side, concerns extending the frontier of criticality to include its out-of-equilibrium properties.
\par Various concepts of dynamical criticality have been studied in classical \cite{Hohenberg1977} and quantum \cite{Taeuber_book} out-of-equilibrium physics. In recent years, \emph{dynamical quantum phase transitions} (DQPT) \cite{Heyl2013} have come under considerable theoretical \cite{Heyl2014, Andraschko2014, Vajna2014, Heyl2015, Vajna2015, Budich2016, Heyl_review} and experimental \cite{Flaeschner2018, Jurcevic2017} investigation. The DQPT concept is based on nonanalyticities in the Loschmidt return rate, a dynamical analog of the equilibrium free energy, where quenches below or above a dynamical critical point lead to different phases characterized by the absence, presence, and nature of nonanalyticities in the return rate. 
\par Though first characterized in free-fermionic two-band models \cite{Heyl2013}, the study of DQPT has been extended to systems with long-range interactions such as transverse-field Ising chains (TFIC) with power-law interactions $\propto1/r^\alpha$, with $r$ interspin distance and $\alpha>0$ \cite{Halimeh2017, Zauner2017, Zunkovic2018}, and in its mean-field limit ($\alpha=0$) \cite{Homrighausen2017,Lang2017,Lang2018}. These studies went beyond the standard DQPT picture \cite{Heyl2013} in which only two kinds of dynamical phases exist: (i) the \emph{trivial} phase (TDP) for quenches before the dynamical critical point where no nonanalyticities (cusps) appear in the return rate; and (ii) the \emph{regular} phase (RDP) where quenching across the dynamical critical point leads to temporally equidistant cusps. Indeed, Ref.~\onlinecite{Halimeh2017} showed that starting in an ordered state in TFIC with sufficiently long-range power-law interactions, the trivial phase is replaced by the so-called \emph{anomalous} dynamical phase (ADP) for sufficiently small quench distance, where, even though one still quenches below the dynamical critical point, cusps arise in the return rate albeit only after its first minimum. As a fundamental characteristic, anomalous cusps are not connected to zero crossings of the order parameter, in contrast to regular cusps when the initial state is ordered \cite{Heyl2014,Halimeh2017,Homrighausen2017}. It was also shown that anomalous cusps belong to a different group of Fisher zeros relative to their regular counterparts~\cite{Halimeh2017,Zauner2017}. ADP was then later found to exist in the integrable limit of full connectedness of TFIC at zero \cite{Homrighausen2017} and finite \cite{Lang2017, Lang2018} temperature. Additionally, ADP seems to coincide with a long-time ordered steady state \cite{Halimeh2017, Zauner2017, Homrighausen2017, Lang2017, Lang2018}. These studies have explored a rich phenomenology of ADP, but leave open whether the origin of ADP lies in the presence of sufficiently long-range interactions, the existence of a finite-temperature phase transition, or yet another physical mechanism.

\begin{figure}[]
\centering
\includegraphics[width=.49\textwidth]{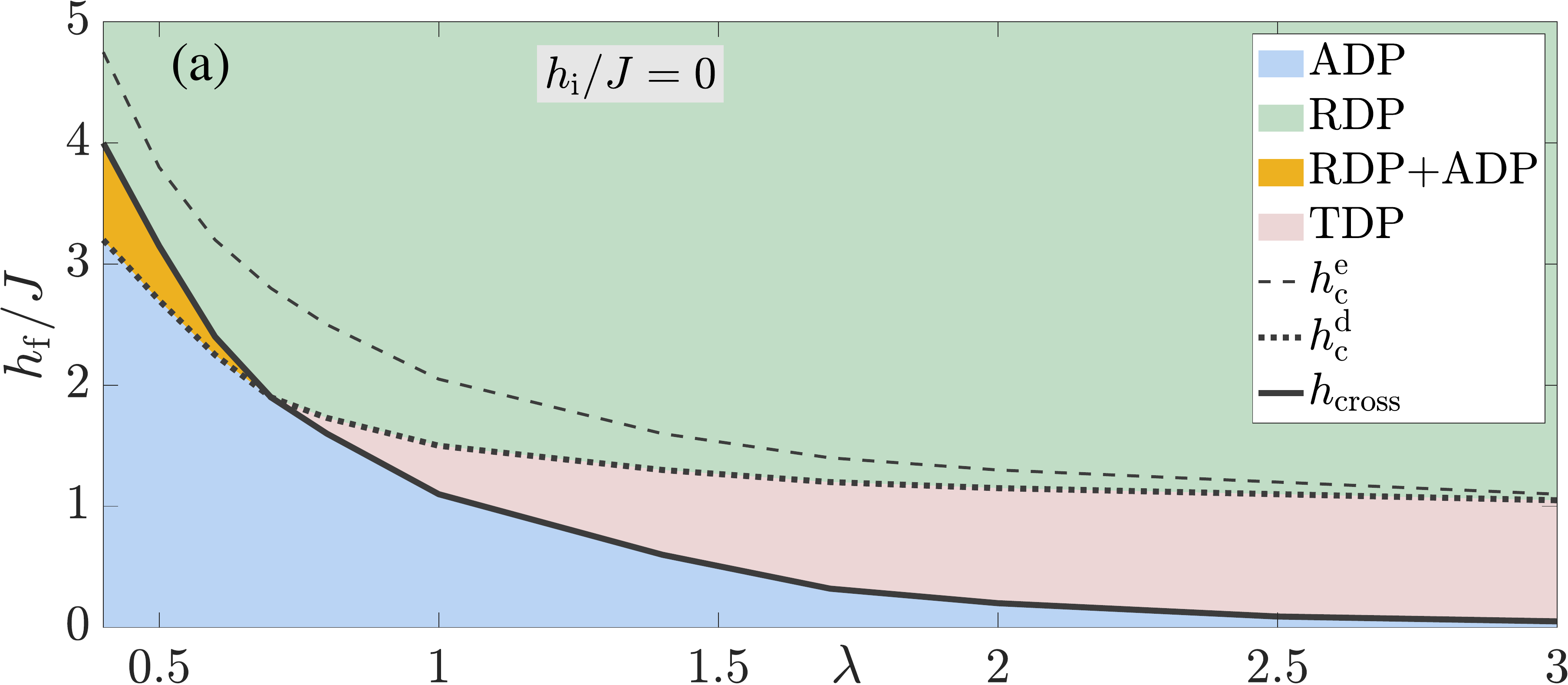}\\
\includegraphics[width=.49\textwidth]{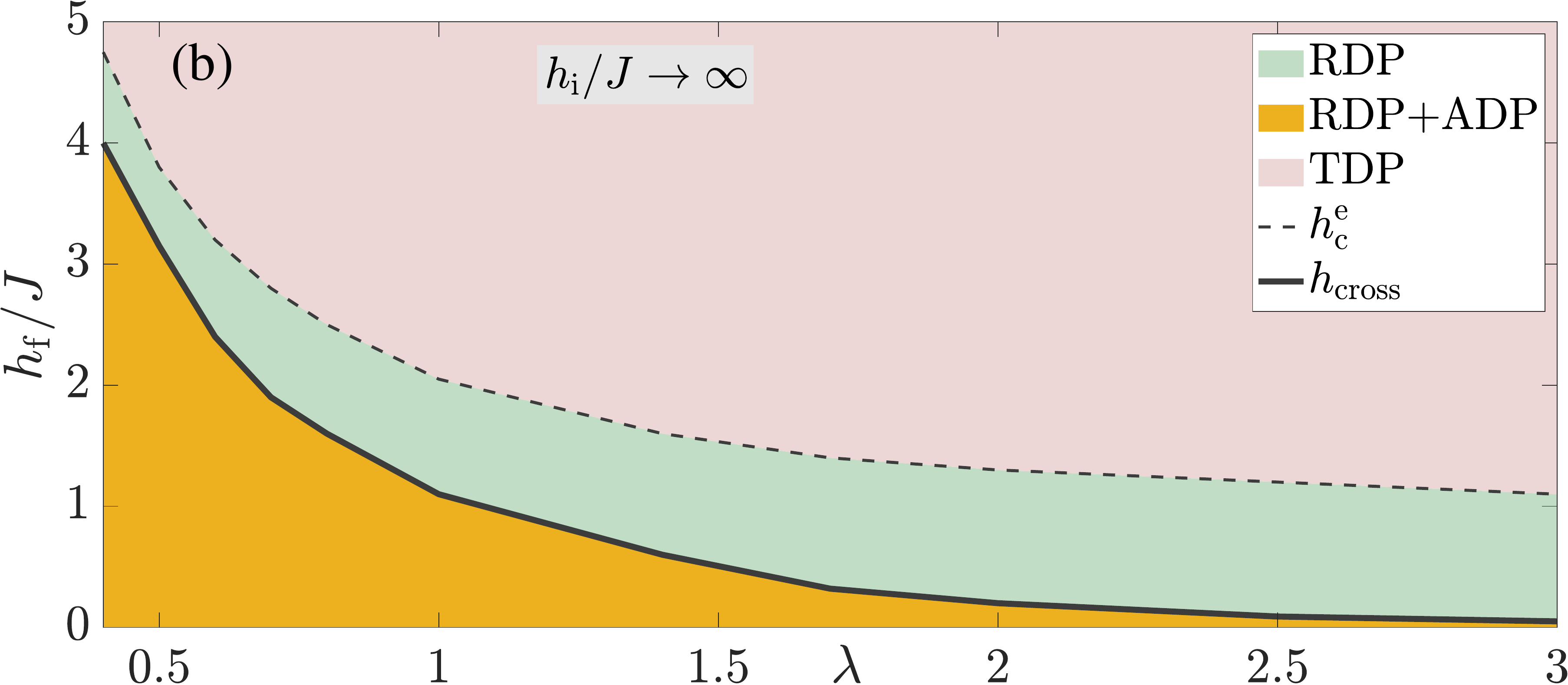}
\caption{(Color online). Dynamical phase diagrams of~\eqref{eq:Ham} in the $(\lambda,h_\mathrm{f})$ plane with an initial transverse-field strength of (a) $h_\mathrm{i}=0$ and (b) $h_\mathrm{i}\to\infty$. The different regions and transitions are explained in the text.}
\label{fig:EPT}
\end{figure}

\par In a seemingly unrelated direction, many efforts have been devoted towards understanding quench dynamics in terms of the ballistic propagation of quasiparticle excitations \cite{Calabrese2006, Calabrese2007a}. These efforts also considered the case of long-range interactions, where the nonlocal nature of the latter can lead to divergences in the quasiparticle group velocity and, consequently, the super-ballistic propagation of information through the system \cite{Hauke2013,Maghrebi2016,Vanderstraeten2018}. In one dimension, long-range interactions can have even more drastic effects on the quasiparticle spectrum: whereas a domain-wall excitation is often the quasiparticle with the lowest energy, long-range interactions across the domain wall lead to a significant increase in its energy and a local excitation can become energetically favorable. In a recent study \cite{Vanderstraeten2018} of the power-law interacting TFIC, it was shown that this scenario leads to a crossover from the `local' regime, where (topologically nontrivial) domain walls are the low-energy quasiparticles, to the `long-range' regime where (topologically trivial) local excitations abound at low energies. This scenario of bound domain walls was recently exploited for observing confined dynamics in long-range interacting spin systems \cite{Liu2018}.
\par In this study, we bring these different directions together for exploring the physical origin of ADP. We provide analytic and numerical evidence that truly long-range interactions are, in fact, not a necessary condition, therefore ruling out a finite-temperature equilibrium phase transition as the origin of ADP. Instead, we present evidence suggesting that the actual origin of ADP is, indeed, the existence of an underlying quasiparticle spectrum crossover.

Our paper is organized as follows: In Sec.~\ref{sec:Model}, we present the model that will form the basis of our numerical results, where the latter are discussed in Sec.~\ref{sec:results}. We review in Sec.~\ref{sec:DW} an analytic proof \cite{Defenu2019,Uhrich2019} showing that domain-wall coupling is a necessary condition for the appearance of anomalous cusps in the Loschmidt return rate. In Sec.~\ref{sec:experiment}, we illustrate how DQPT can be used as an experimental probe to map out the equilibrium physics of the underlying model. We conclude in Sec.~\ref{sec:conclusion}, and include additional supporting analytic arguments in Appendix~\ref{sec:Landau}, numerical results at different parameter values in Appendix~\ref{sec:Larger}, and convergence evidence in Appendix~\ref{sec:Convergence}.

\section{Model}\label{sec:Model}
In previous works, TFIC with power-law decaying interactions has been studied, and the existence of ADP has been established \cite{Halimeh2017, Zauner2017}. In this paper, we study the case of TFIC with exponentially decaying interactions, given by the Hamiltonian
\begin{align}\label{eq:Ham}
\hat{H}(h)= -J \sum_{j>l} \mathrm{e}^{-\lambda(|l-j|-1)}\hat{\sigma}_l^z\hat{\sigma}_j^z-h\sum_{l=1}^N\hat{\sigma}_l^x,
\end{align}
where $\hat{\sigma}_j^{\{x,y,z\}}$ are the Pauli matrices on site $j$, $J>0$ is the spin-coupling constant, $h$ is the transverse-field strength, and $\lambda>0$. The model exhibits a quantum phase transition from a symmetry-broken ground state (small $h$) to a polarized state (large $h$), where the critical field $h_\mathrm{c}^\mathrm{e}$ shifts as $\lambda$ decreases (see Fig.~\ref{fig:EPT}). The physics of the exponentially decaying interactions can be adiabatically connected to the nearest-neighbor case ($\lambda\to\infty$), and we expect that the phase transition is in the same universality class. Moreover, this excludes a finite-temperature phase transition, as is confirmed by a simple Landau-Lifshitz argument \cite{Landau2013,Thouless1969}; see Appendix~\ref{sec:Landau}.

\begin{figure}[]
\centering
\includegraphics [width=0.48\textwidth]{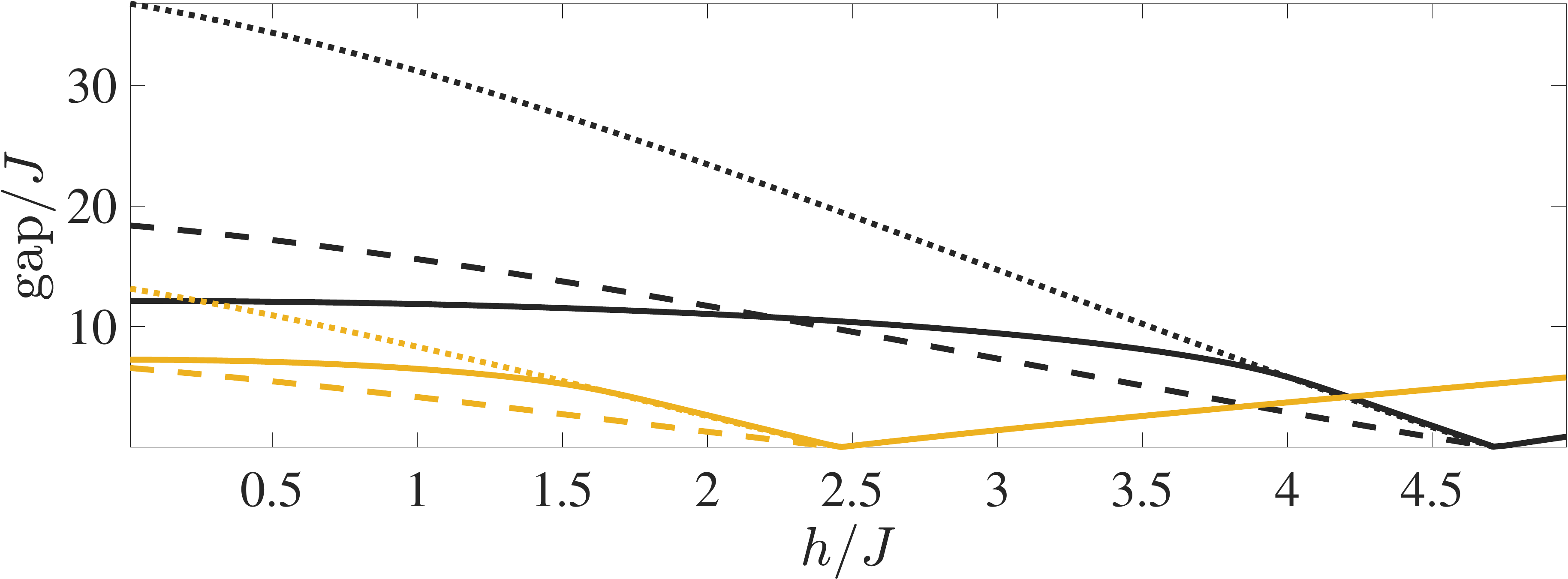}
\caption{(Color online). The energy gap in the topologically nontrivial (dashed lines) and topologically trivial (solid lines) sectors for~\eqref{eq:Ham} $\lambda=0.4$ (black/dark) and $\lambda=0.8$ (yellow/light) as function of transverse-field strength $h$. The dotted line indicates the edge of the two-domain-wall continuum: if we find an excitation below this continuum, this corresponds to a stable local excitation. Both gaps approach zero at the equilibrium critical point. There is no topological sector in the paramagnetic phase, so we cannot define a crossover there.
}
\label{fig:crossover}
\end{figure}

\begin{figure}[htp]
\centering
\includegraphics[width=.49\textwidth]{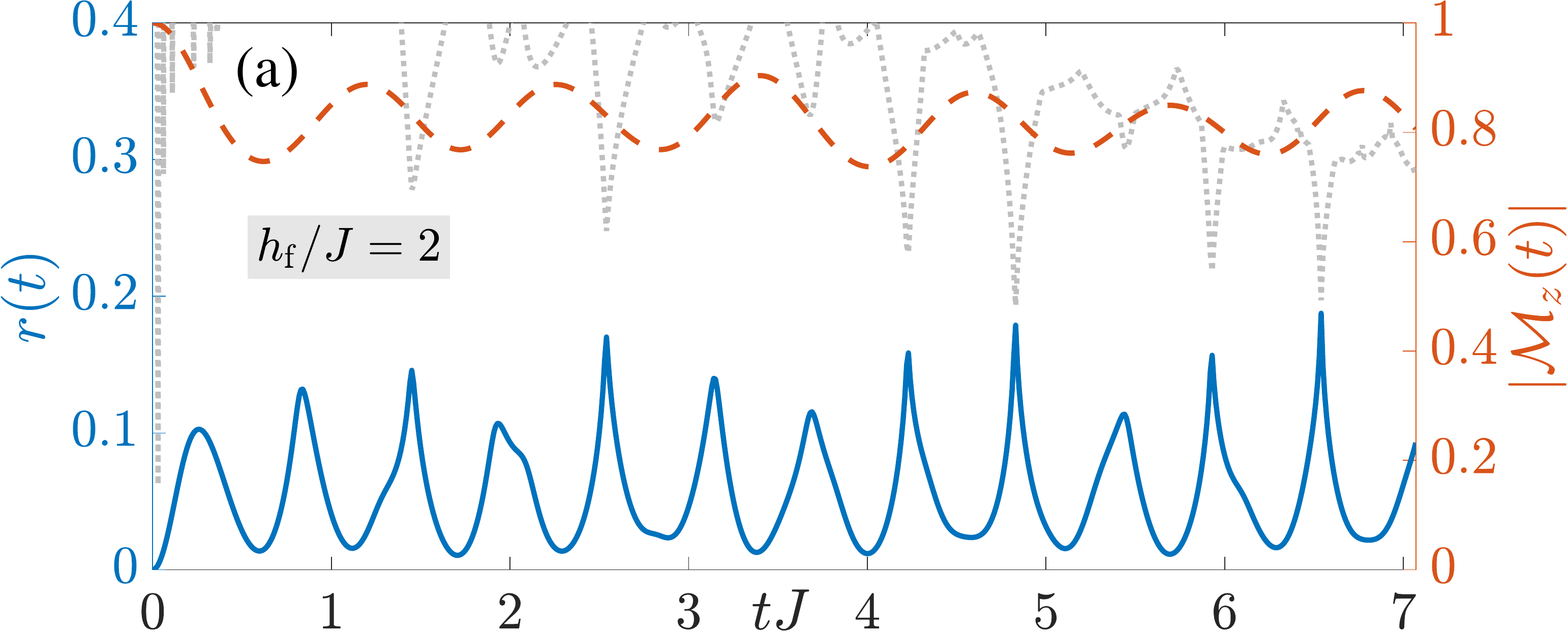}\\
\includegraphics[width=.49\textwidth]{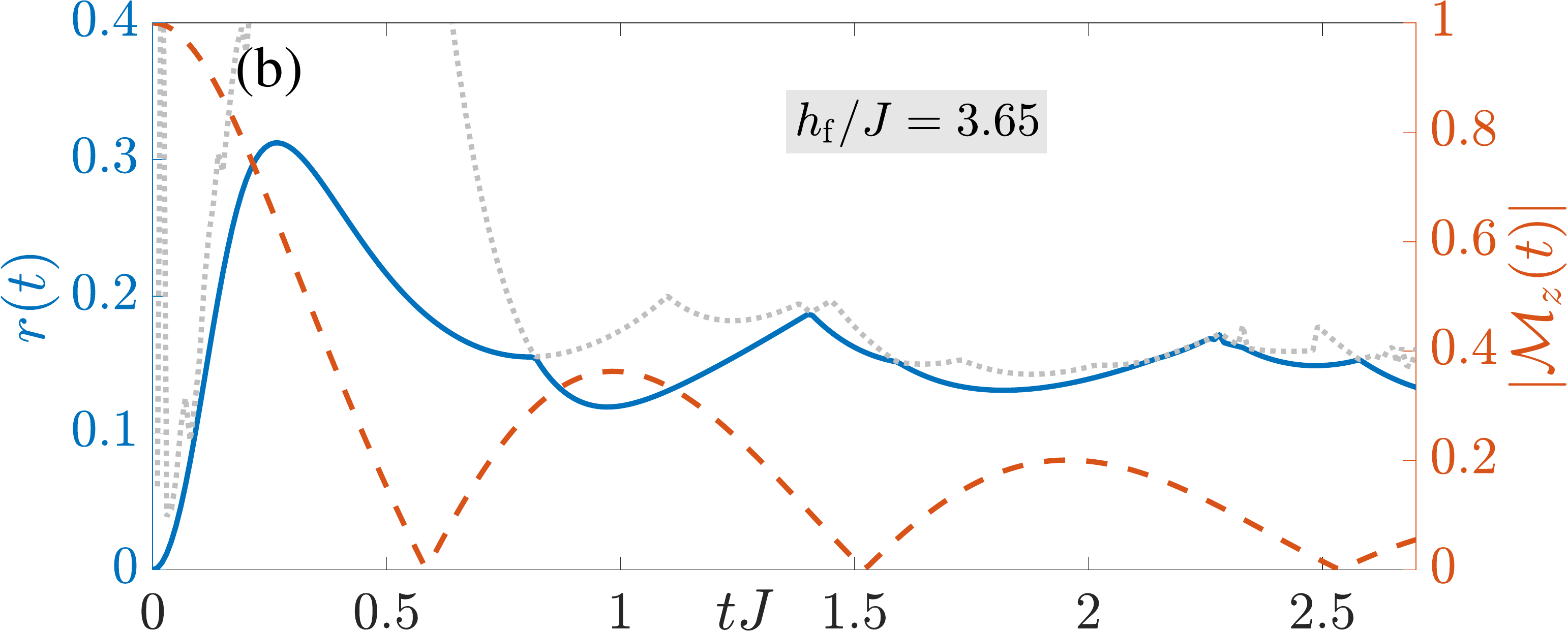}\\
\includegraphics[width=.49\textwidth]{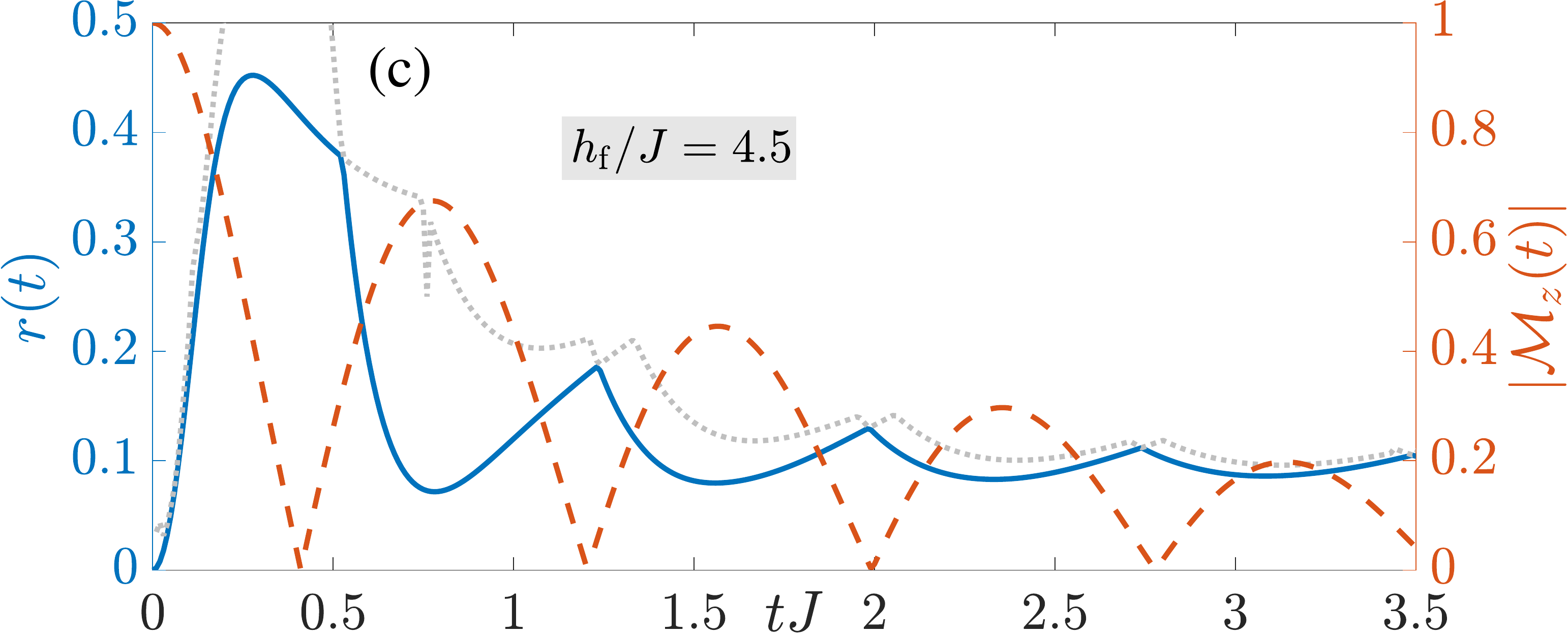}
\caption{(Color online). Loschmidt return rates after quenching a fully $z$-polarized state with $\hat{H}(h_\mathrm{f})$ for $\lambda=0.4$ at (a) $h_\mathrm{f}=2J$, (b) $h_\mathrm{f}=3.65J$, and (c) $h_\mathrm{f}=4.5J$. Dotted gray lines indicate sections of the branch cut above the return rate (solid blue line), and cusps form when they intersect, whereas the dashed orange line is the order parameter. Here two phases appear separately as ADP for small quenches $h_\mathrm{f}<h_\mathrm{c}^\mathrm{d}<h_\mathrm{cross}$ as seen in (a), and RDP for large quenches $h_\mathrm{f}>h_\mathrm{cross}>h_\mathrm{c}^\mathrm{d}$ as displayed in (b), but the return rate also exhibits both cusps for intermediate quenches $h_\mathrm{f}\in(h_\mathrm{c}^\mathrm{d},h_\mathrm{cross})$ as shown in (c). Note that for $\lambda=0.4$ and quenches from $h_\mathrm{i}=0$ we have $h_\mathrm{c}^\mathrm{d}<h_\mathrm{cross}$ [see Fig.~\ref{fig:EPT}(a)]. See Fig.~\ref{fig:ZQuenchExplam0p8} where at $\lambda=0.8$ and quenches from $h_\mathrm{i}=0$ we have $h_\mathrm{c}^\mathrm{d}>h_\mathrm{cross}$.}
\label{fig:ZQuenchExplam0p4} 
\end{figure}

In the short-range model (large $\lambda$), the symmetry-broken phase hosts domain-wall excitations that interpolate between the two ferromagnetic ground states. Upon decreasing $\lambda$, the domain walls become more massive (because of the interactions between different ground-state configurations across the domain walls) and a local excitation goes down in energy. In the limit $h=0$, these two different types of quasiparticles can be understood, on the one hand, as a bare domain wall and, on the other, a single spin flip on one of the two ground states. In this limit, their energies can be derived as
\begin{align}\label{eq:topoCost}
\mathcal{E}_\text{dw}&=\lim_{N\to\infty}\sum_{l,r=1}^{N/2}\text{e}^{-\lambda(r+l-2)}=\frac{\text{e}^\lambda}{2(\cosh\lambda-1)},\\\label{eq:trivCost}
\mathcal{E}_\text{local}&=\lim_{N\to\infty}2\sum_{r=1}^{N/2} \text{e}^{-\lambda (r-1)}=\frac{2}{1-\text{e}^{-\lambda}},
\end{align}
respectively. A crossover between these two excitations at $h=0$ occurs at $\lambda_\text{c}=\log 2\approx0.69$. In the presence of a magnetic field ($h>0$), the quasiparticles will become dressed by quantum fluctuations and these energies will start to shift, crossing at $\lambda<\lambda_\text{c}$. We have used the matrix product state (MPS) quasiparticle Ansatz \cite{Vanderstraeten2018,Vanderstraeten2019} to compute the excitation gaps in the two sectors in the quantum regime that confirm this picture, see Fig.~\ref{fig:crossover}: In the `local' regime the lowest-lying excitation in the topologically trivial sector is a two-domain-wall scattering state, whereas in the `long-range' regime there is a stable local excitation that is below the two-domain-wall continuum. We define $h_\text{cross}(\lambda)$ as the value of the transverse-field strength separating the two regions; cf.~Fig.~\ref{fig:EPT}.

\begin{figure}[htp]
\centering
\includegraphics[width=.49\textwidth]{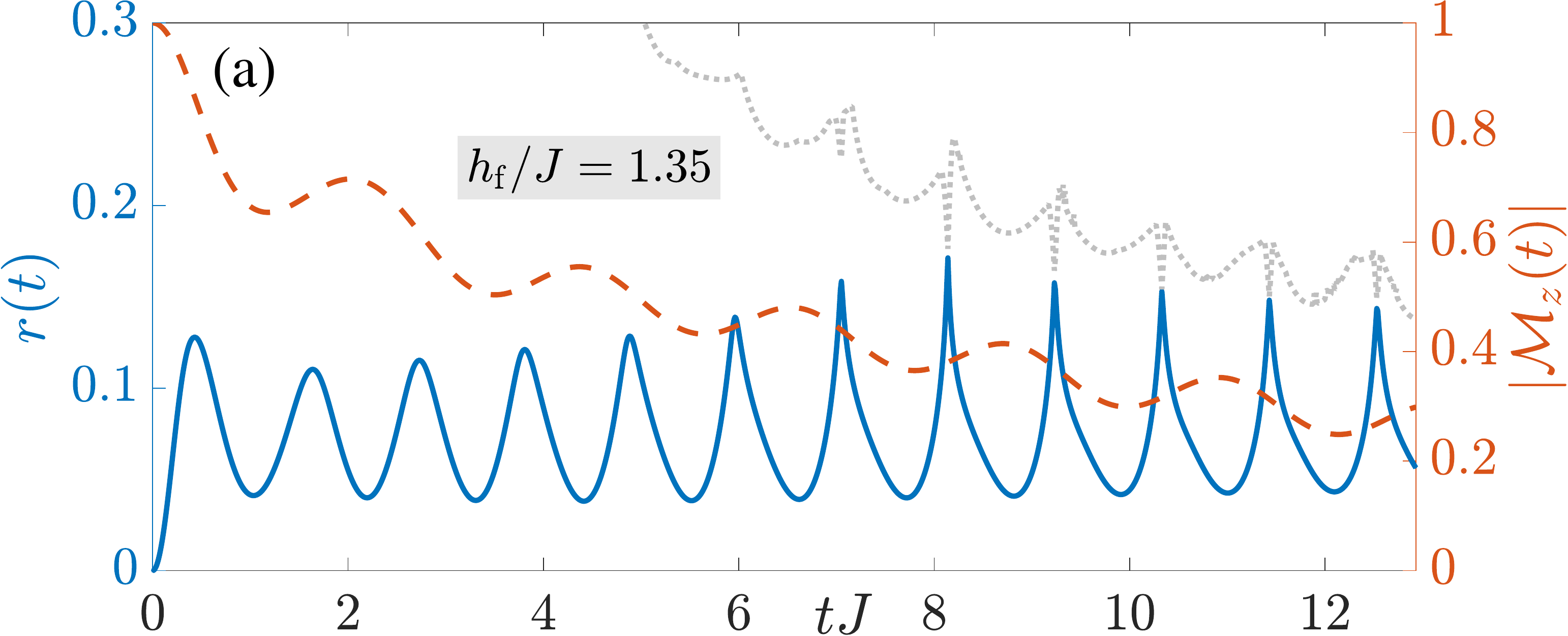}\\
\includegraphics[width=.49\textwidth]{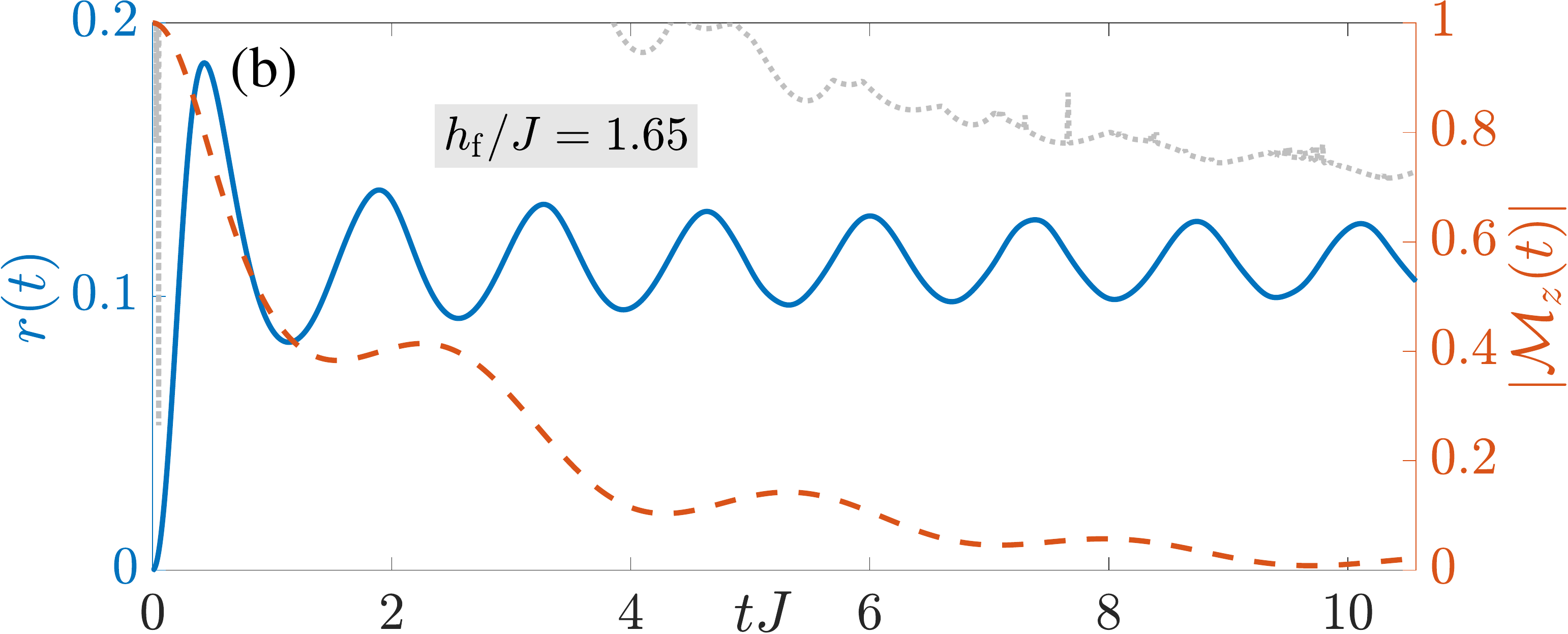}\\
\includegraphics[width=.49\textwidth]{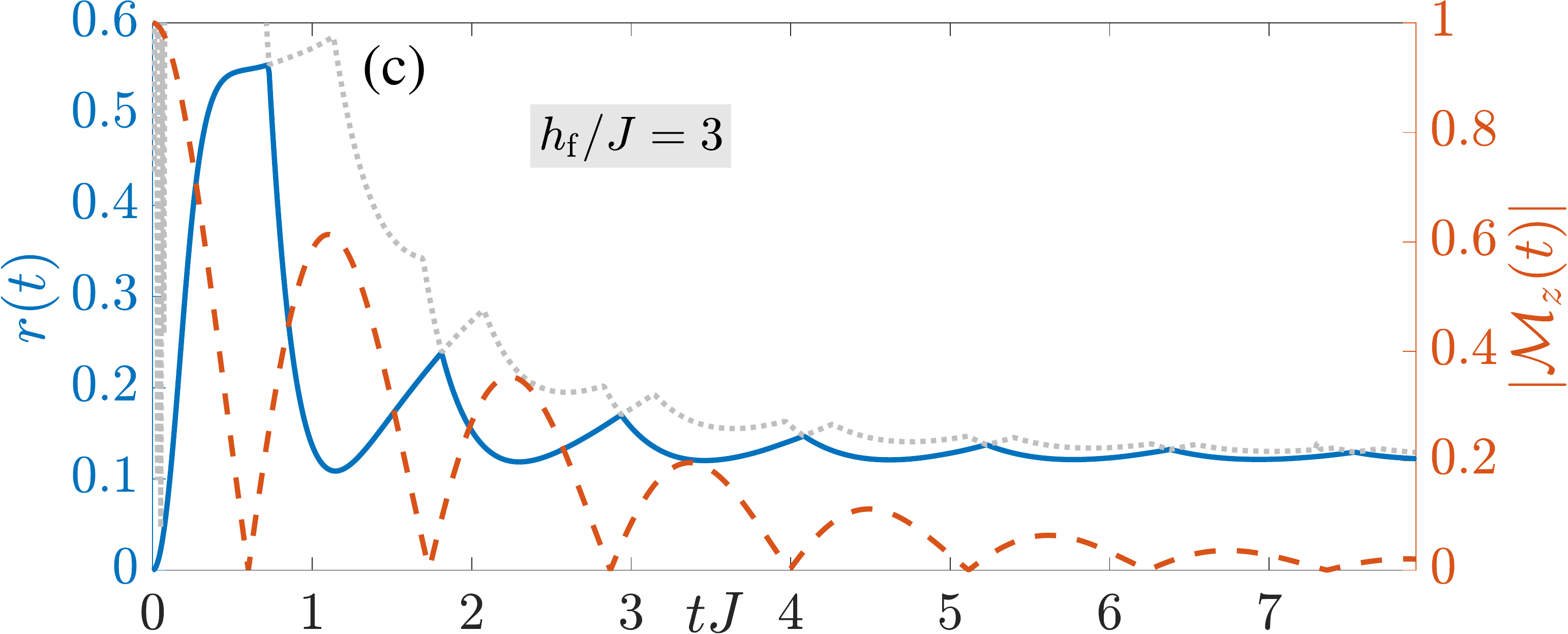}
\caption{(Color online). Same as Fig.~\ref{fig:ZQuenchExplam0p4} but for $\lambda=0.8$ with (a) $h_\mathrm{f}=1.35J$, (b) $h_\mathrm{f}=1.65J$, and (c) $h_\mathrm{f}=3J$. The return rate shows one of three distinct phases: ADP for small quenches $h_\mathrm{f}<h_\mathrm{cross}<h_\mathrm{c}^\mathrm{d}$ as shown in (a), TDP for intermediate quenches $h_\mathrm{f}\in(h_\mathrm{cross},h_\mathrm{c}^\mathrm{d})$ as seen in (b), and RDP for large quenches $h_\mathrm{f}>h_\mathrm{c}^\mathrm{d}>h_\mathrm{cross}$ as displayed in (c). Note that for $\lambda=0.8$ and quenches from $h_\mathrm{i}=0$ we have $h_\mathrm{c}^\mathrm{d}>h_\mathrm{cross}$ [see Fig.~\ref{fig:EPT}(a)]. See Fig.~\ref{fig:ZQuenchExplam0p4} where at $\lambda=0.4$ and quenches from $h_\mathrm{i}=0$ we have $h_\mathrm{c}^\mathrm{d}<h_\mathrm{cross}$.}
\label{fig:ZQuenchExplam0p8} 
\end{figure}

\begin{figure}[htp]
\centering
\includegraphics[width=.49\textwidth]{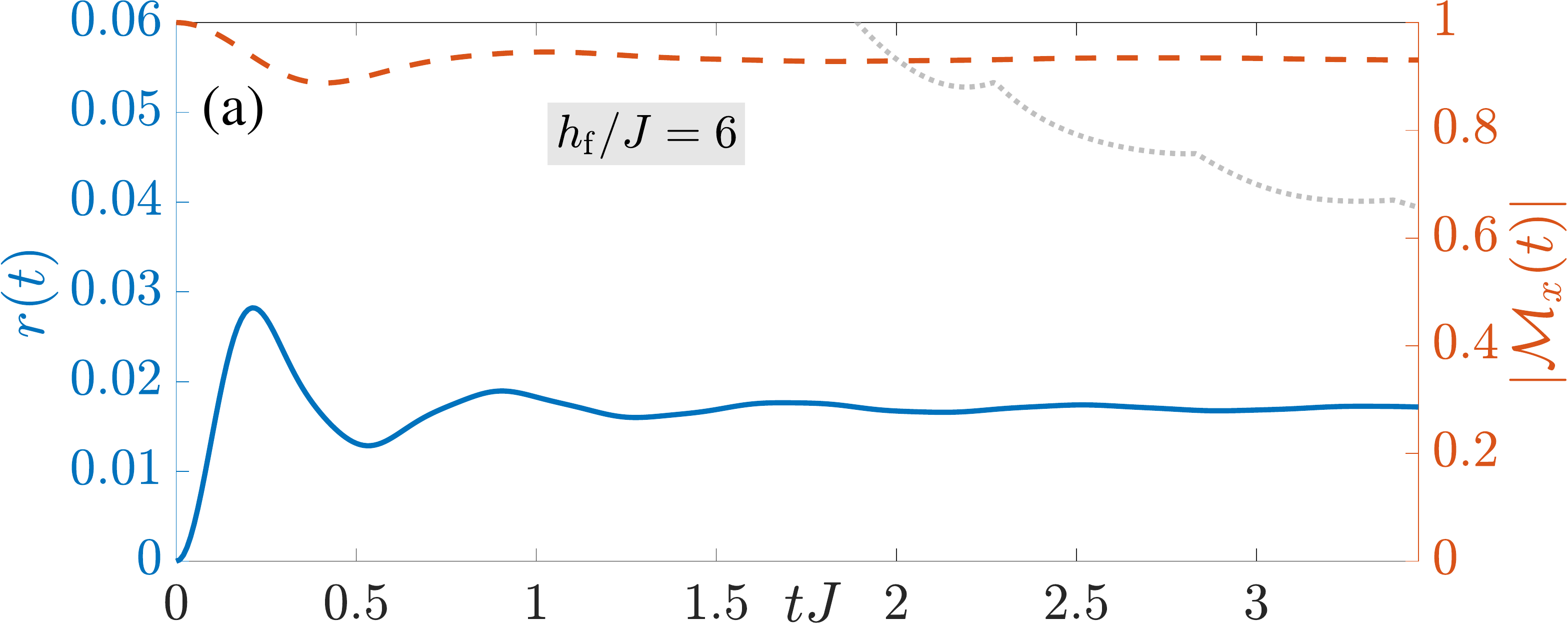}\\
\includegraphics[width=.49\textwidth]{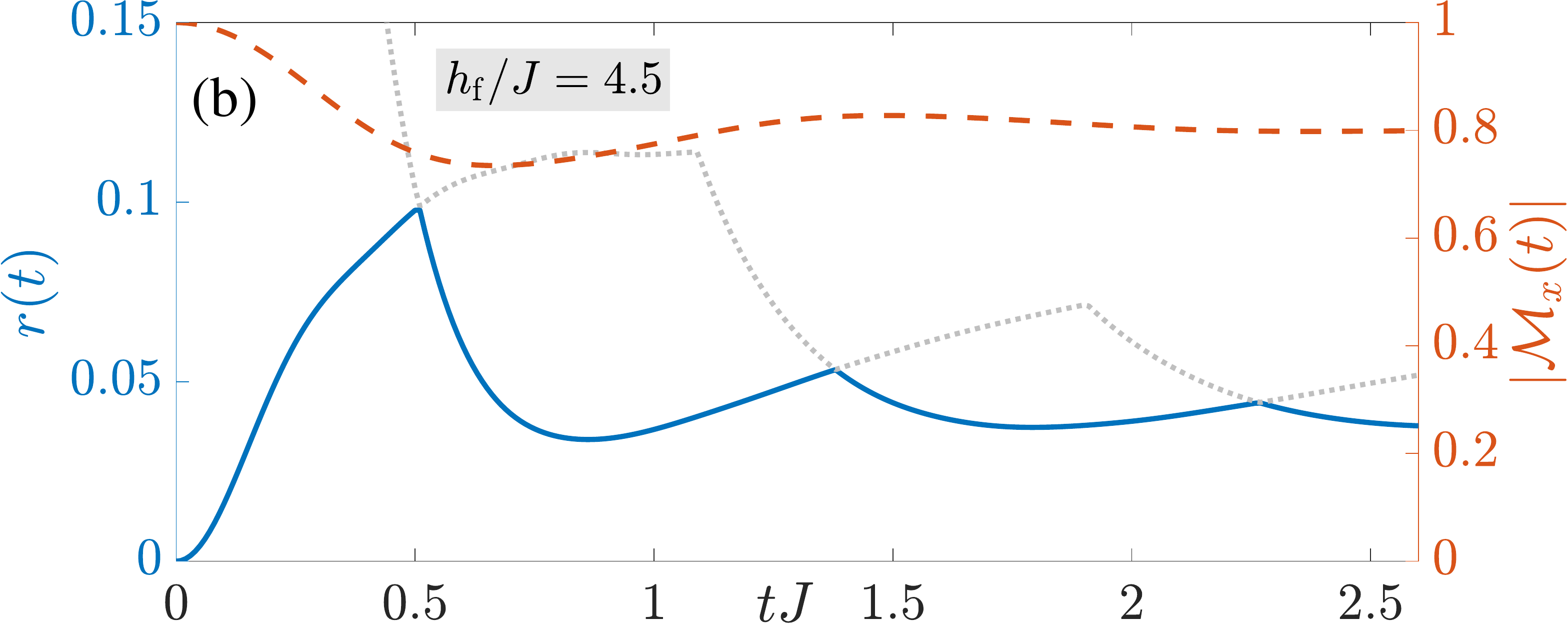}\\
\includegraphics[width=.49\textwidth]{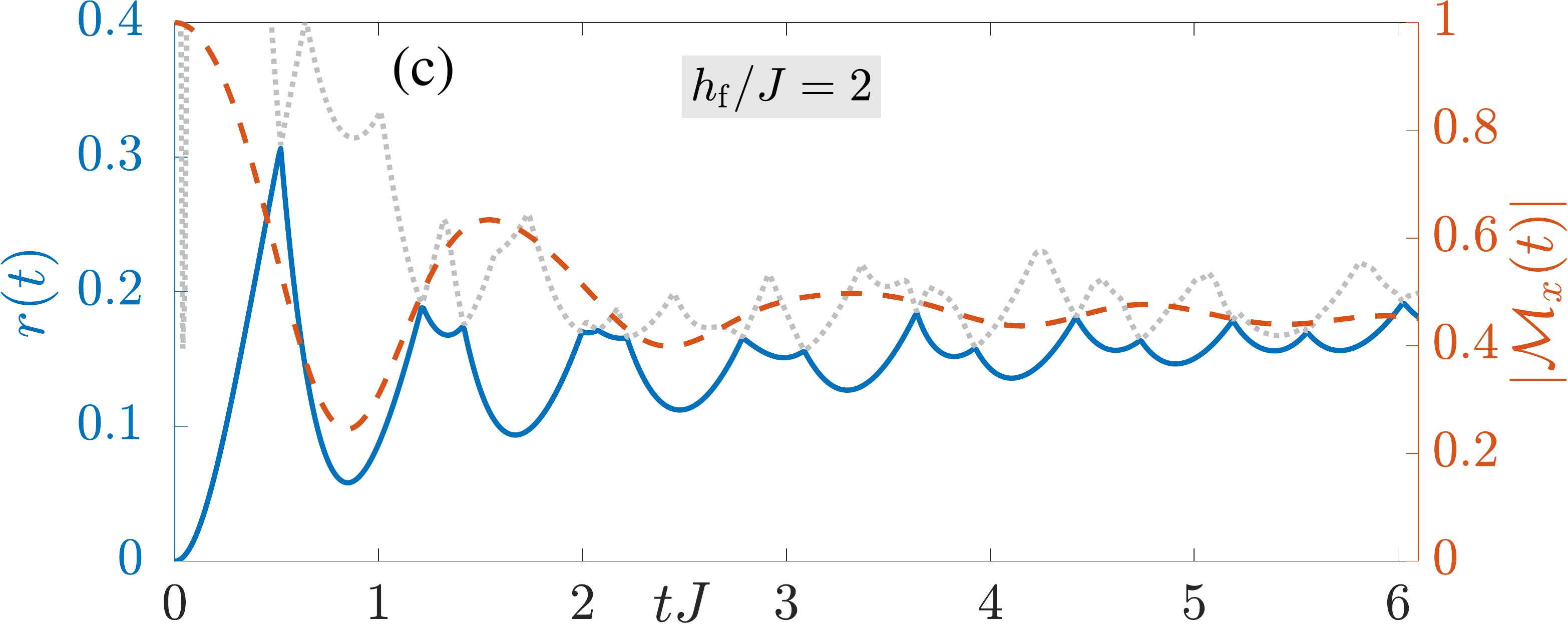}
\caption{(Color online). Same as Fig.~\ref{fig:ZQuenchExplam0p4} but starting from a fully $x$-polarized state ($h_\mathrm{i}\to\infty$), and while showing the transverse magnetization $\mathcal{M}_x(t)$ rather than the order parameter, as the latter is always zero here due to both the initial state and quenching Hamiltonian being $\mathrm{Z}_2$-symmetric. (a) The return rate shows TDP for $h_\mathrm{f}=6J>h_\mathrm{c}^\mathrm{e}$, (b) RDP for $h_\mathrm{f}=4.5J\in(h_\mathrm{cross},h_\mathrm{c}^\mathrm{e})$, and (c) both regular and anomalous cusps for $h_\mathrm{f}=2J<h_\mathrm{cross}$.}
\label{fig:XQuenchExplam0p4} 
\end{figure}

\begin{figure}[htp]
\centering
\includegraphics[width=.49\textwidth]{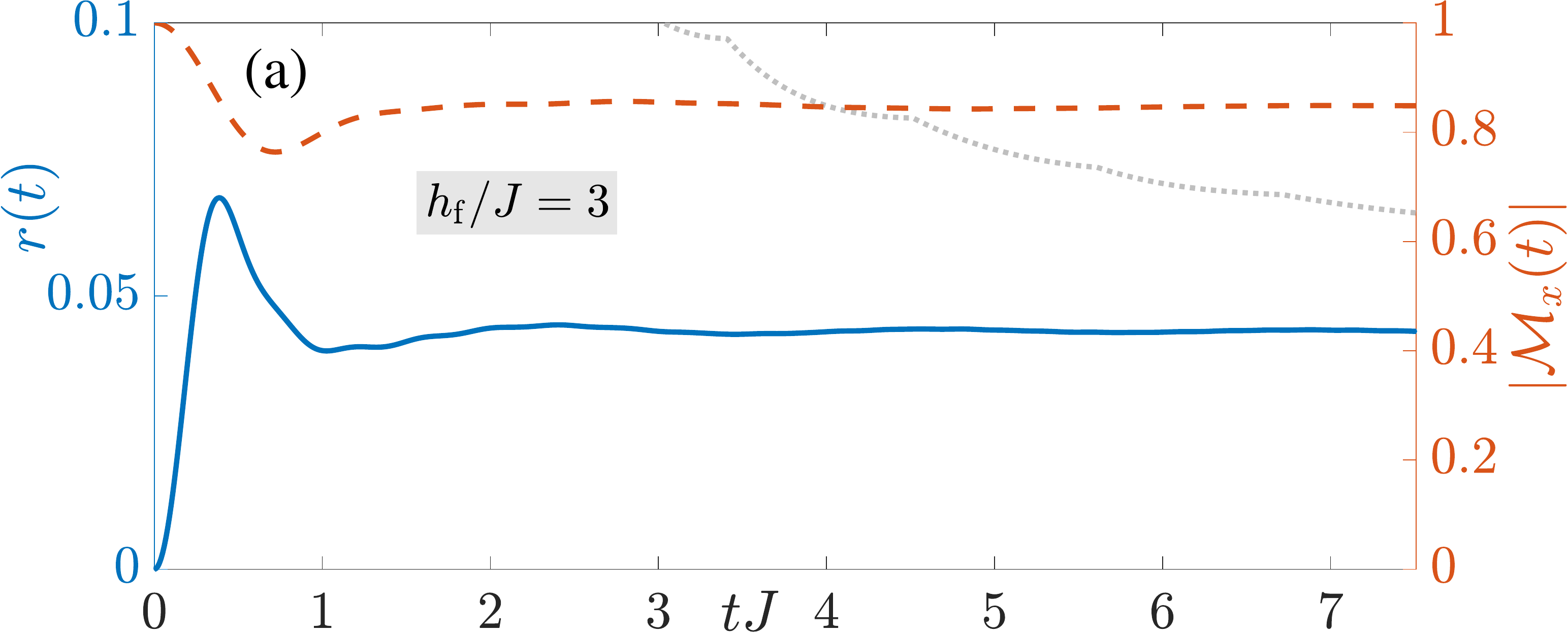}\\
\includegraphics[width=.49\textwidth]{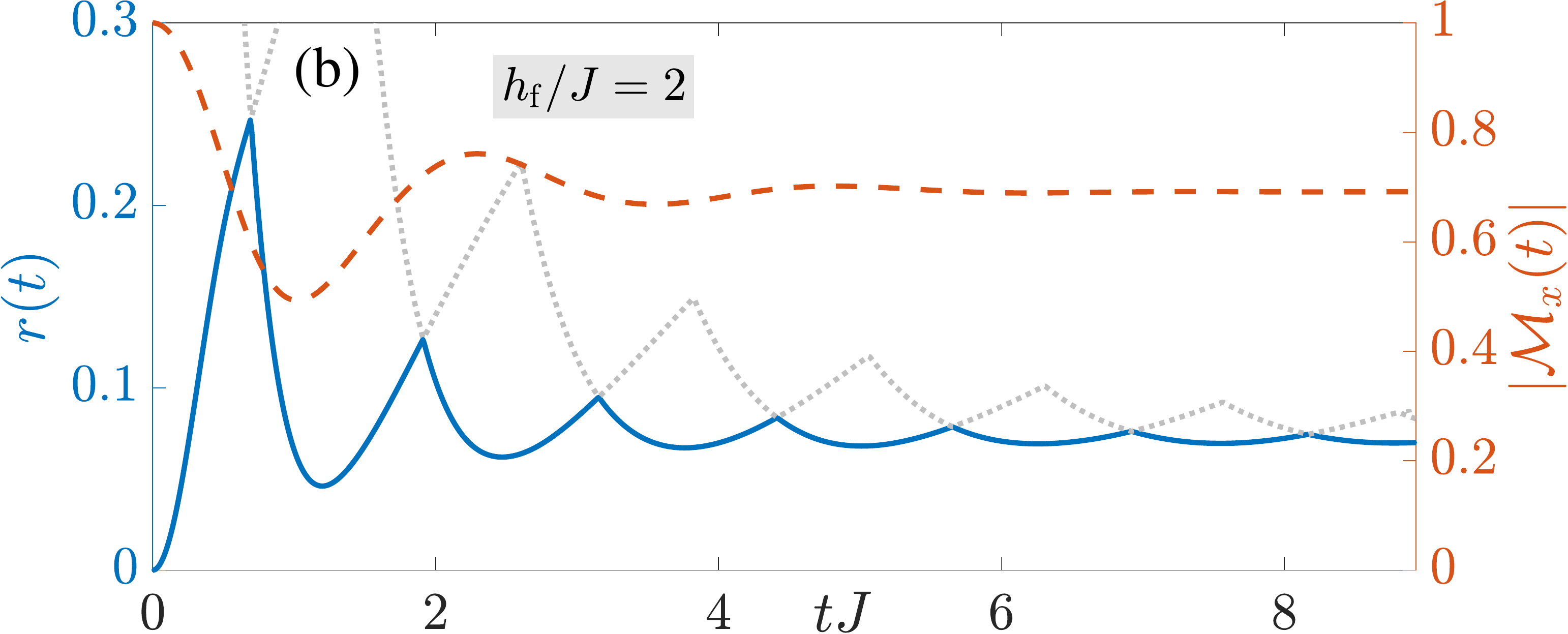}\\
\includegraphics[width=.49\textwidth]{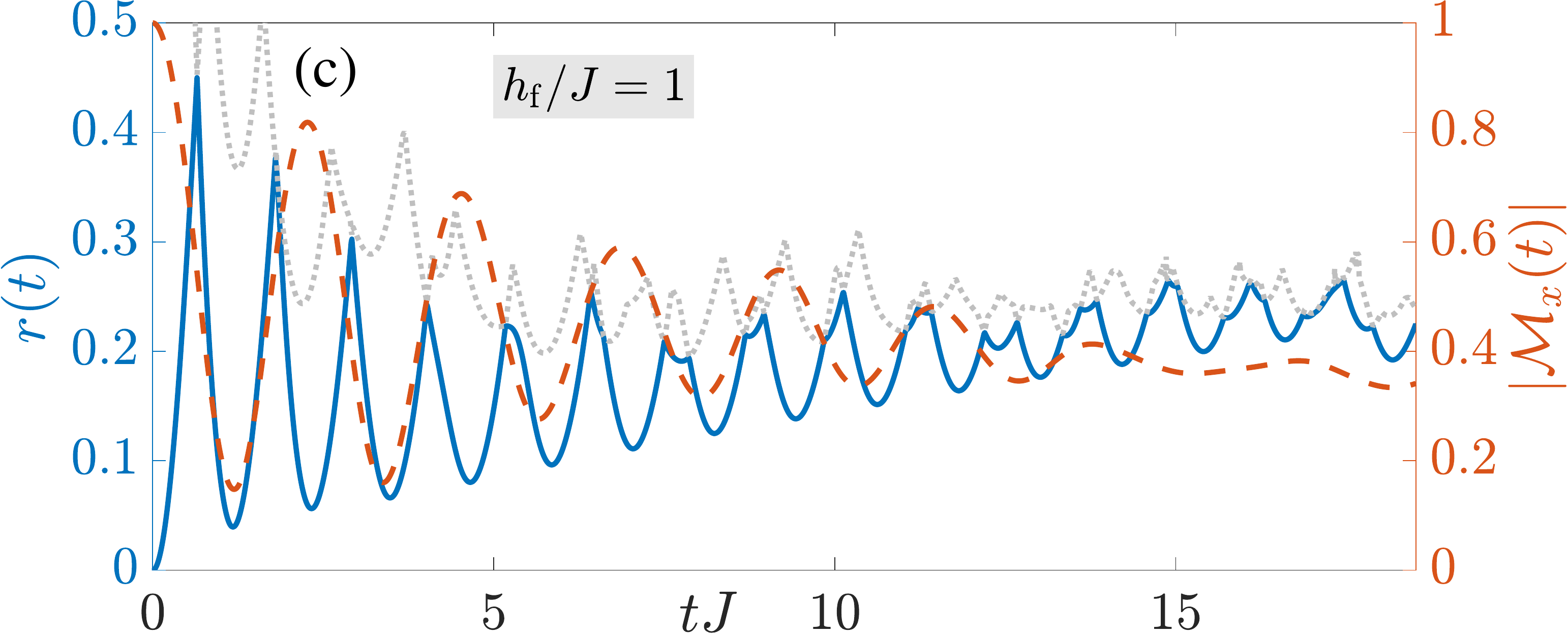}
\caption{(Color online). Same as Fig.~\ref{fig:XQuenchExplam0p4} but for $\lambda=0.8$ with (a) $h_\mathrm{f}=3J>h_\mathrm{c}^\mathrm{e}$, (b) $h_\mathrm{f}=2J\in(h_\mathrm{cross},h_\mathrm{c}^\mathrm{e})$, and (c) $h_\mathrm{f}=J<h_\mathrm{cross}$. The dynamical behavior is qualitatively the same as at other values $\lambda$ for this quench.}
\label{fig:XQuenchExplam0p8} 
\end{figure}

\section{Results and discussion}\label{sec:results}
We now present our numerical results, where we simulate the quench dynamics using uniform MPS and the time-dependent variational principle \cite{Haegeman2011, Haegeman2016,Vanderstraeten2019}. The Loschmidt return rate is defined as
\begin{align}\label{eq:RR}
r(t)=-\lim_{N\to\infty}\frac{1}{N}\ln\big|\langle\psi_\mathrm{i}|\text{e}^{-\text{i}\hat{H}(h_\mathrm{f})t}|\psi_\mathrm{i}\rangle\big|^2,
\end{align}
with $|\psi_\mathrm{i}\rangle$ the ground state of $\hat{H}(h_\mathrm{i})$. The return rate can be computed in MPS from the return-rate branches, which are the (negative of the) logarithms of the eigenvalues of the mixed MPS transfer matrix \cite{Zauner2017}. Nonanalyticities in the return rate emerge when the two lowest-lying branches intersect, thereby making the detection of cusps straightforward. Simple observables such as 
\begin{align}\label{eq:OP}
\mathcal{M}_{\{x,z\}}(t)=\lim_{N\to\infty}\frac{1}{N}\sum_{j=1}^N\langle\hat{\sigma}_j^{\{x,z\}}(t/2)\rangle,
\end{align}
are readily evaluated using MPS. Here, $\mathcal{M}_z(t)$ is the Landau order parameter. Post-quench dynamics of the order parameter have been studied in the nearest-neighbor \cite{Calabrese2011,Calabrese2012} and power-law interacting \cite{Halimeh2017b} TFIC, the XXZ chain \cite{Barmettler2009,Heyl2014}, and the Bose-Hubbard model \cite{Altman2002}. For quenches from an ordered initial state, the order parameter makes zero crossings (changes sign) only for quenches across a dynamical critical point, while asymptotically going to zero (as an envelope) in all cases in the absence of a finite-temperature phase transition.
In our model, which is nonintegrable and in the short-range universality class (see Appendix~\ref{sec:Landau}), $\mathcal{M}_z(t)$ is therefore expected to go to zero for all quenches, but to make zero crossings only for quenches above a dynamical critical point $h_\mathrm{c}^\mathrm{d}$ defined on this basis.% Moreover, these zero crossings have been shown to correspond to regular cusps in the return rate for Ising-like models \cite{Heyl2014,Halimeh2017}.

Let us prepare our system in the fully $z$-polarized state, a ground state of $\hat{H}(h_\mathrm{i}=0)$, and quench with $\hat{H}(h_\mathrm{f})$. First we look at the case of $\lambda=0.4$ shown in Fig.~\ref{fig:ZQuenchExplam0p4}. For quenches below the dynamical critical point $h_\mathrm{c}^\mathrm{d}$, the return rate displays anomalous cusps that do not correspond to any zero crossings in the order parameter. For quenches to $h_\mathrm{f}\in(h_\mathrm{c}^\mathrm{d},h_\mathrm{cross})$ we observe (regular) cusps that are directly connected to zero crossings in $\mathcal{M}_z(t)$ in addition to (anomalous) cusps that are not, signifying a sort of coexistence of both ADP and RDP. Our results indicate that this `coexistence region' of ADP and RDP happens only when $h_\mathrm{cross}>h_\text{f}>h_\mathrm{c}^\mathrm{d}$, which can occur only for sufficiently small $\lambda\lesssim\lambda_\text{c}$. On the other hand, when $h_\mathrm{f}>h_\mathrm{cross}$, only regular cusps exist, showing the same periodicity in time as zero crossings in the order parameter. The picture qualitatively changes for $\lambda=0.8$ shown in Fig.~\ref{fig:ZQuenchExplam0p8}. Here $h_\mathrm{cross}<h_\mathrm{c}^\mathrm{d}$, which leads to anomalous cusps for $h_\mathrm{f}<h_\mathrm{cross}$, regular cusps for $h_\mathrm{f}>h_\text{c}^\text{d}$, and, interestingly, a smooth return rate for $h_\mathrm{cross}<h_\mathrm{f}<h_\mathrm{c}^\mathrm{d}$. Indeed, the return rate shows a cusp in its eighth peak at $h_\mathrm{f}=1.35J$, which then smoothens out at $h_\mathrm{f}=1.65J$, contrary to the property of cusp proliferation with increasing $h_\mathrm{f}$ when ADP borders RDP \cite{Halimeh2017,Homrighausen2017,Lang2017}.

These results suggest that the occurrence of cusps in the return rate is connected to the stability of local excitations in the system. Indeed, in the local regime where freely propagating domain-wall excitations dominate, we observe the dynamical properties of the nearest-neighbor case, but in the long-range regime the domain walls are bound into a stable local excitation, and only then ADP emerges. This connection implies that ADP \emph{always} exists in our model for finite positive $\lambda$, albeit it shrinks ($h_\text{cross}$ gets smaller) with increasing $\lambda$ and completely disappears for $\lambda\to\infty$.

In order to further confirm this picture, we repeat the above quench procedure but starting in the fully $x$-polarized state, the ground state of $\hat{H}(h_\mathrm{i}\to\infty)$. For the case of $\lambda=0.4$ in Fig.~\ref{fig:XQuenchExplam0p4}, the return rate shows cusps only for quenches across the equilibrium critical point $h_\mathrm{c}^\mathrm{e}$, in agreement with the nearest-neighbor limit. However, even though only regular cusps appear for $h_\mathrm{cross}<h_\mathrm{f}<h_\mathrm{c}^\mathrm{e}$ that are evenly spaced in time, for larger quenches $h_\mathrm{f}<h_\mathrm{cross}$ we observe both anomalous cusps, which are unevenly spaced in time, and regular cusps. The results for $\lambda=0.8$ in Fig.~\ref{fig:XQuenchExplam0p8} and larger $\lambda$ values are qualitatively the same (see Appendix~\ref{sec:Larger} for results at larger $\lambda$ values). It is worth noting that even though the first cusps may appear regular for quenches to $h_\mathrm{f}<h_\mathrm{cross}$, the higher branch-cut segments are qualitatively different from the case of $h_\mathrm{cross}<h_\mathrm{f}<h_\mathrm{c}^\mathrm{d}$. In Figs.~\ref{fig:XQuenchExplam0p4} and~\ref{fig:XQuenchExplam0p8} we also show $\mathcal{M}_x(t)$. We see roughly a common periodicity between the inflection points of this observable and the regular cusps, although not much can be deduced from this, because $\mathcal{M}_x(t)$ is not the order parameter. The latter is here always zero because both $|\psi_\mathrm{i}\rangle$ and $\hat{H}(h_\mathrm{f})$ possess $\mathrm{Z}_2$ symmetry.

Finally, we note that in our numerical simulations we have used a maximum bond dimension $D_\mathrm{max}=350$ and a time-step $\delta t=10^{-3}/J$, at which convergence is achieved for all our results. Since we work in the thermodynamic limit directly, no finite-size errors are present; see Appendix~\ref{sec:Convergence} for further details.

\section{Domain-wall coupling as a necessary condition for anomalous cusps}\label{sec:DW}
We shall now provide an analytic argument as to why domain-wall coupling is a necessary condition for anomalous cusps to appear in the Loschmidt return rate \cite{Defenu2019,Uhrich2019}. Starting with the Hamiltonian~\eqref{eq:Ham}, we employ the Jordan-Wigner transformation

\begin{align}
&\hat{\sigma}^x_j=1-2\hat{c}_j^\dagger \hat{c}_j,\\
&\hat{\sigma}^y_j=-\mathrm{i}\left[\prod_{n=1}^{j-1}(1-2\hat{c}_n^\dagger \hat{c}_n)\right](\hat{c}_j-\hat{c}_j^\dagger),\\
&\hat{\sigma}^z_j=-\left[\prod_{n=1}^{j-1}(1-2\hat{c}_n^\dagger \hat{c}_n)\right](\hat{c}_j+\hat{c}_j^\dagger),
\end{align}
which renders~\eqref{eq:Ham} in the form

\begin{align}\nonumber
\hat{H}(h)=&\,-J \sum_{j>l} \mathrm{e}^{-\lambda(|l-j|-1)}(\hat{c}_l^\dagger-\hat{c}_l)\\\nonumber
&\times\left[\prod_{n=l+1}^{j-1}(1-2\hat{c}_n^\dagger \hat{c}_n)\right](\hat{c}_j+\hat{c}_j^\dagger)\\\label{eq:HamJW}
&-h\sum_{l=1}^N(1-2\hat{c}_l^\dagger \hat{c}_l),
\end{align}
where $\hat{c}_l,\hat{c}_l^\dagger$ are the fermionic annihilation and creation operators, respectively, on site $l$ satisfying the canonical anticommutation relations $\{\hat{c}_l,\hat{c}_j\}=0$ and $\{\hat{c}_l,\hat{c}_j^\dagger\}=\delta_{l,j}$.

The Hamiltonian~\eqref{eq:HamJW} contains higher-than-quadratic terms in the fermionic operators, which renders it unsolvable exactly. Therefore, we simplify it into the quadratic form 

\begin{align}\nonumber
\hat{H}(h)=&\,-J \sum_{j>l} \mathrm{e}^{-\lambda(|l-j|-1)}(\hat{c}_l^\dagger\hat{c}_j-\hat{c}_l\hat{c}_j+\hat{c}_l^\dagger\hat{c}_j^\dagger-\hat{c}_l\hat{c}_j^\dagger)\\\label{eq:HamTJW}
&-h\sum_{l=1}^N(1-2\hat{c}_l^\dagger \hat{c}_l),
\end{align}
by employing the approximation

\begin{align}\label{eq:TJW}
\prod_{n=l+1}^{j-1}(1-2\hat{c}_n^\dagger \hat{c}_n)=1,\,\,\,\forall j>l+1.
\end{align}
The Hamiltonian~\eqref{eq:HamTJW} never gives rise to anomalous cusps in the return rate, but rather shows regular cusps for quenches across its topological equilibrium critical point \cite{Dutta2017,Defenu2019,Uhrich2019}. It is known, however, that the Jordan-Wigner fermion operator $\hat{c}_j=\hat{\sigma}_j^x(\hat{\sigma}_j^z-\mathrm{i}\hat{\sigma}_j^y)$ is equivalent to a domain-wall creation operator in the spin picture \cite{Fradkin_book}. As such, the quartic terms in the Jordan-Wigner fermionic operators eliminated in the approximation~\eqref{eq:TJW} represent interactions between domain walls in the spin picture. Consequently, this shows that domain-wall coupling is a necessary condition for the appearance of anomalous cusps in the Loschmidt return rate. Intriguingly, in the limit of $\lambda\to\infty$, when the Hamiltonian~\eqref{eq:Ham} is that of the nearest-neighbor quantum Ising chain, the quench dynamics can never lead to anomalous cusps according to the conclusions of our main text. This agrees with the analytic argument we make here since in the nearest-neighbor limit the Hamiltonians~\eqref{eq:HamJW} and~\eqref{eq:HamTJW} are identical.

\section{DQPT as an experimental probe of equilibrium physics}\label{sec:experiment}
Determining equilibrium universality from short-time quench dynamics \cite{Karl2017} is a very attractive prospect given the limited evolution times that are accessible in modern experiments. Based on our results, it becomes clear that DQPT can be used as an experimental probe to estimate the equilibrium properties of the underlying model. Indeed, our results show that quenches from the fully disordered state of the quantum Ising model give rise to return rates with nonanalyticities directly indicative of the equilibrium quantum critical point and also the crossover value of the transverse-field strength below which domain walls are bound. In particular, we see that a final value of the transverse-field strength equal to the equilibrium critical point separates between the trivial and regular dynamical phases. On the other hand, when the final value of the transverse-field strength is equal to its crossover value, this separates between the regular dynamical phase and the coexistence region. Consequently, these different nonanalyticities, which occur at relatively short times, can be an efficient means in ultracold-atom and ion-trap setup to unravel equilibrium physics, especially when in such experiments long evolution times are a major challenge. It is to be noted that even though we have validated this conclusion in the particular case of the transverse-field Ising chain with exponentially decaying interactions, we expect our results to apply to more general models such as the transverse-field Ising chain with power-law interactions, and, in fact, have recently been validated in the two-dimensional quantum Ising model \cite{Hashizume2018}.

\section{Conclusion}\label{sec:conclusion}
We have provided numerical evidence linking the existence of a quasiparticle spectrum crossover between local and two-domain-wall excitations in the topologically trivial quasiparticle sector to anomalous criticality in the return rate, which for quenches within the ordered phase does not correspond to any zero crossings in the order parameter. This is demonstrated in the transverse-field Ising chain with exponentially decaying interactions, where anomalous criticality arises regardless of the initial state, only disappearing in the integrable nearest-neighbor limit. As a consequence, our results show that models in the same equilibrium universality class can host drastically different out-of-equilibrium properties: for any finite positive $\lambda$, the dynamical phase diagram is qualitatively different from that of the nearest-neighbor quantum Ising chain. Moreover, our study resolves the outstanding question as to whether anomalous cusps are associated with truly long-range interactions or a finite-temperature phase transition, as here we observe anomalous cusps in a short-range model that has neither. 
\par We expect our picture to be valid for generic systems with crossovers between different types of quasiparticles \cite{Hashizume2018}. Indeed, our results already explain the appearance of anomalous cusps for small quenches within the ferromagnetic phase of the fully connected quantum Ising model \cite{Lang2017,Lang2018}, which only hosts local excitations in that phase. Similarly, our conclusions have been validated in two-dimensional Ising square \cite{Hashizume2018} and triangular \cite{Hashizume2019} lattices where quasiparticles are always local excitations due to the unbounded cost of domain walls in $d>1$ dimensions. Importantly, our results should be experimentally accessible in modern ultracold-atom \cite{Flaeschner2018} and ion-trap \cite{Jurcevic2017} setups, which have already detected regular cusps. 

\section*{Acknowledgments}
The authors gratefully acknowledge stimulating discussions with Bernhard Frank, Markus Heyl, Johannes Lang, Achilleas Lazarides, Francesco Piazza, and Matthias Punk. This work is supported by the Research Foundation Flanders, ERC grants QUTE (647905) and ERQUAF (715861).
\bibliography{OriginADP_biblio}

\appendix

\begin{figure*}[htp]
	\centering
	\includegraphics[width=.49\textwidth]{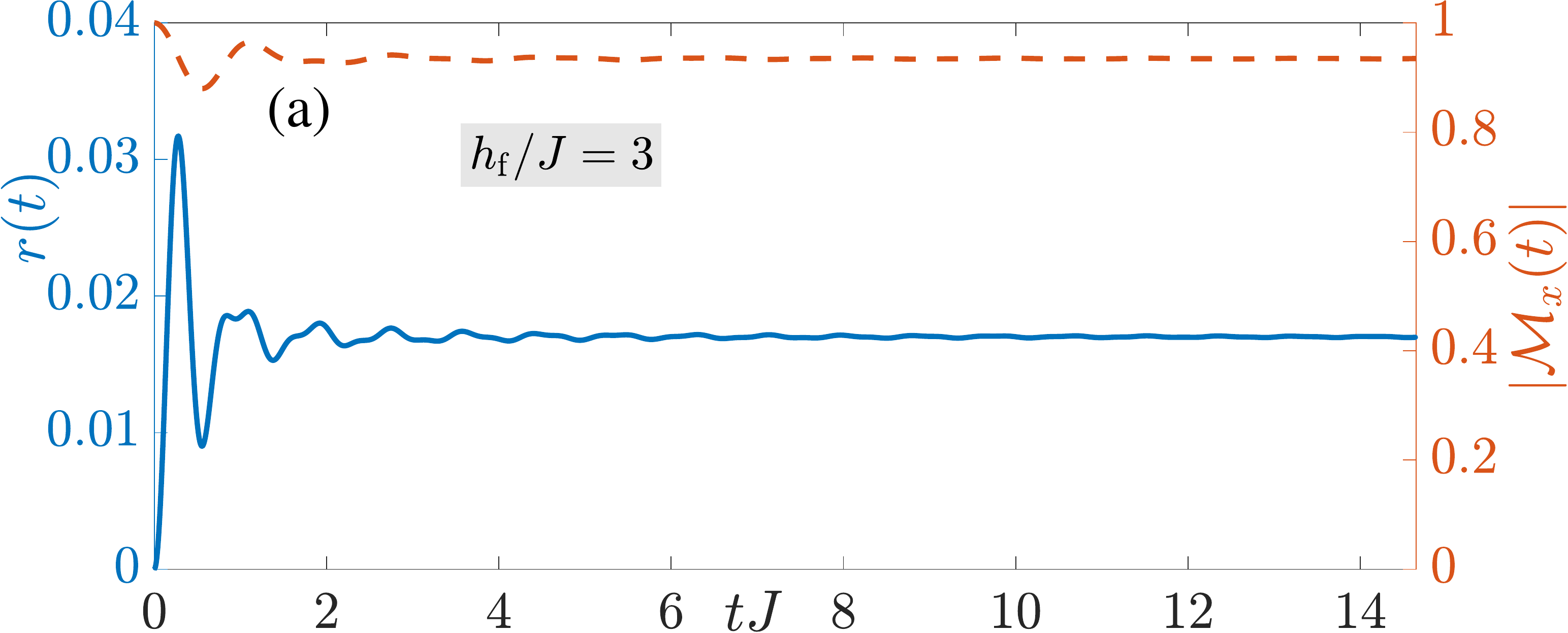}\quad
	\includegraphics[width=.49\textwidth]{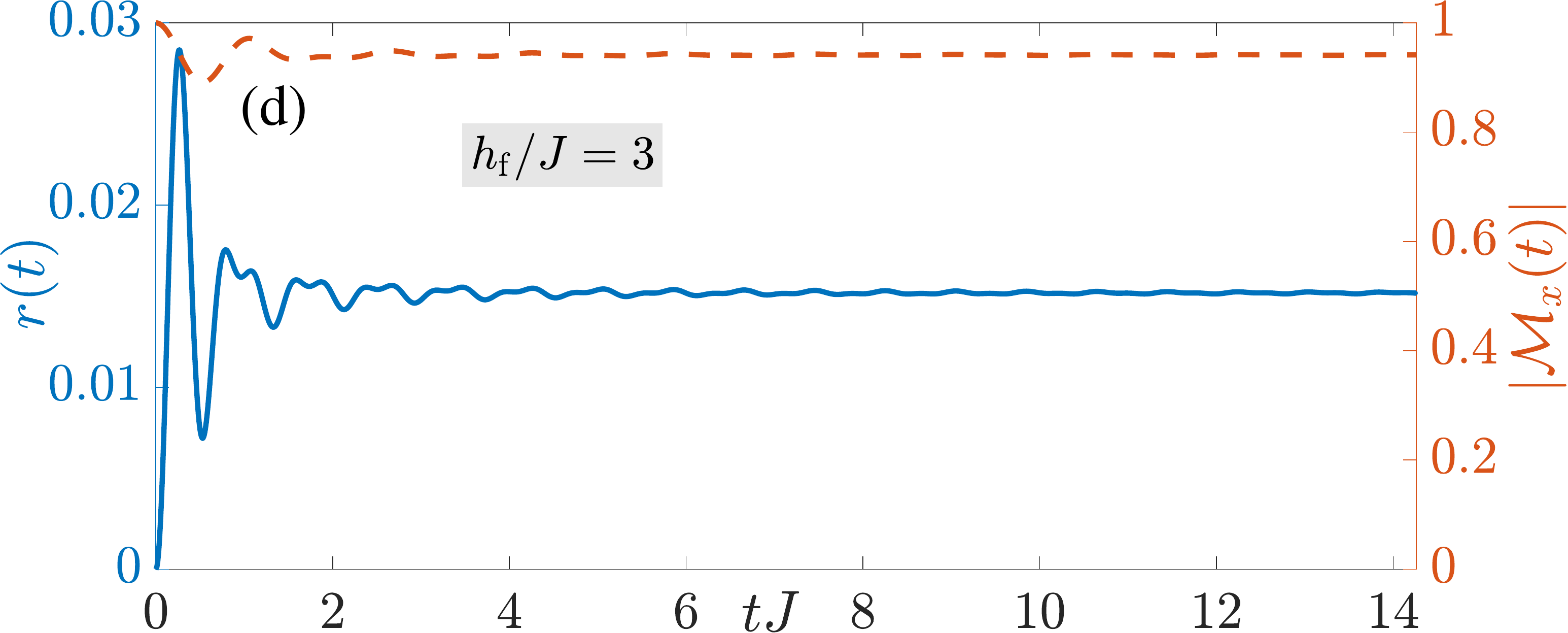}\quad
	\includegraphics[width=.49\textwidth]{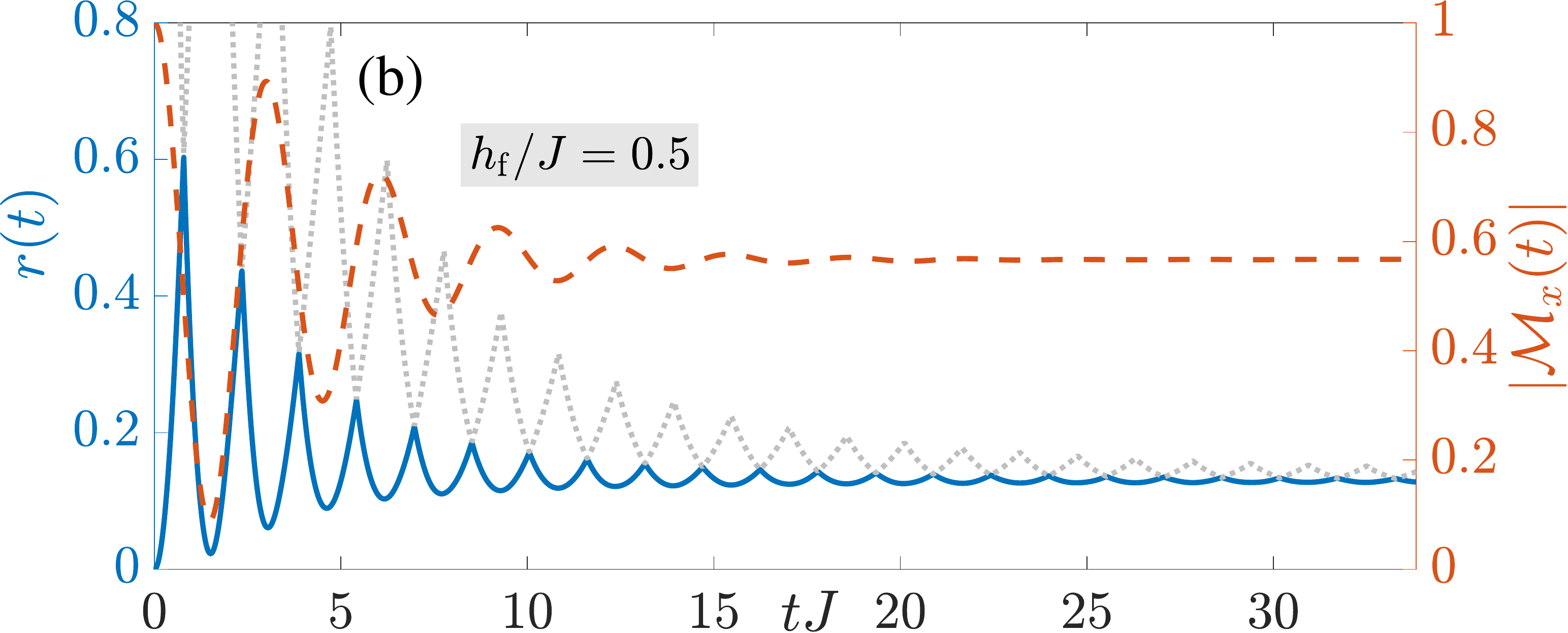}\quad
	\includegraphics[width=.49\textwidth]{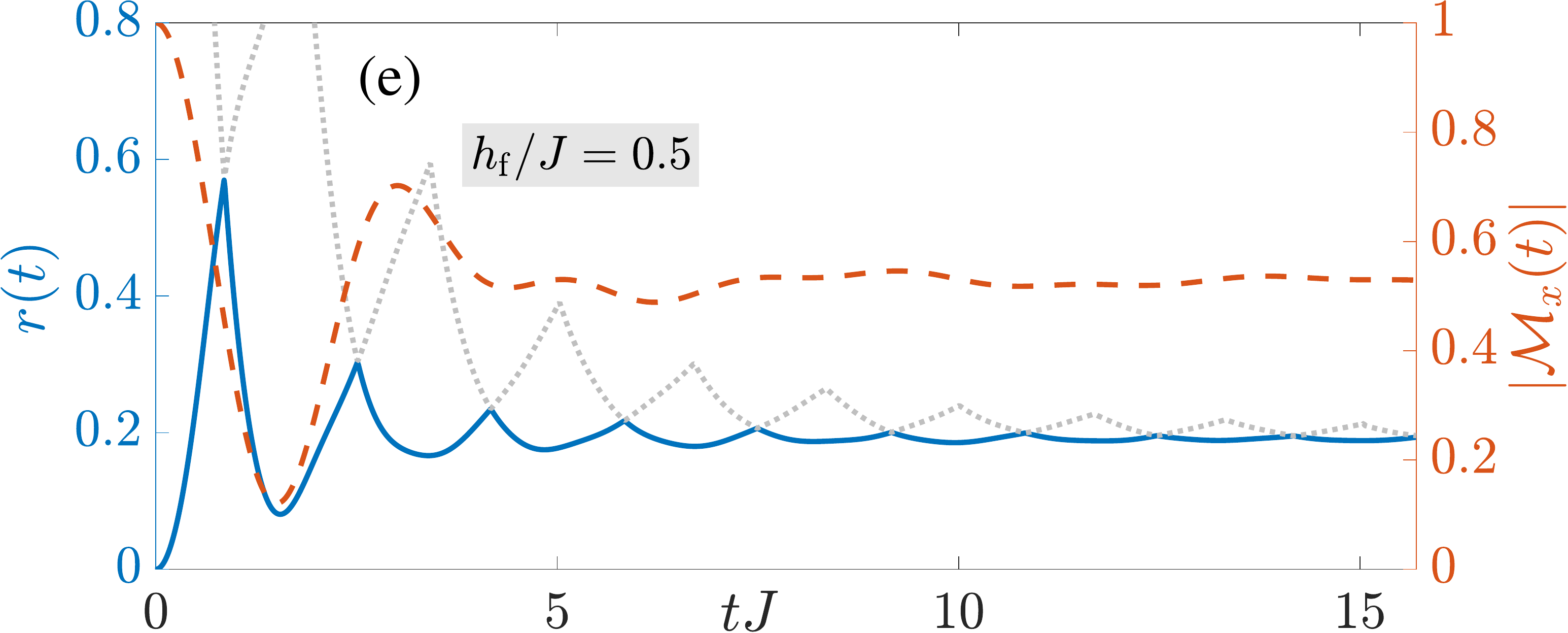}\\
	\includegraphics[width=.49\textwidth]{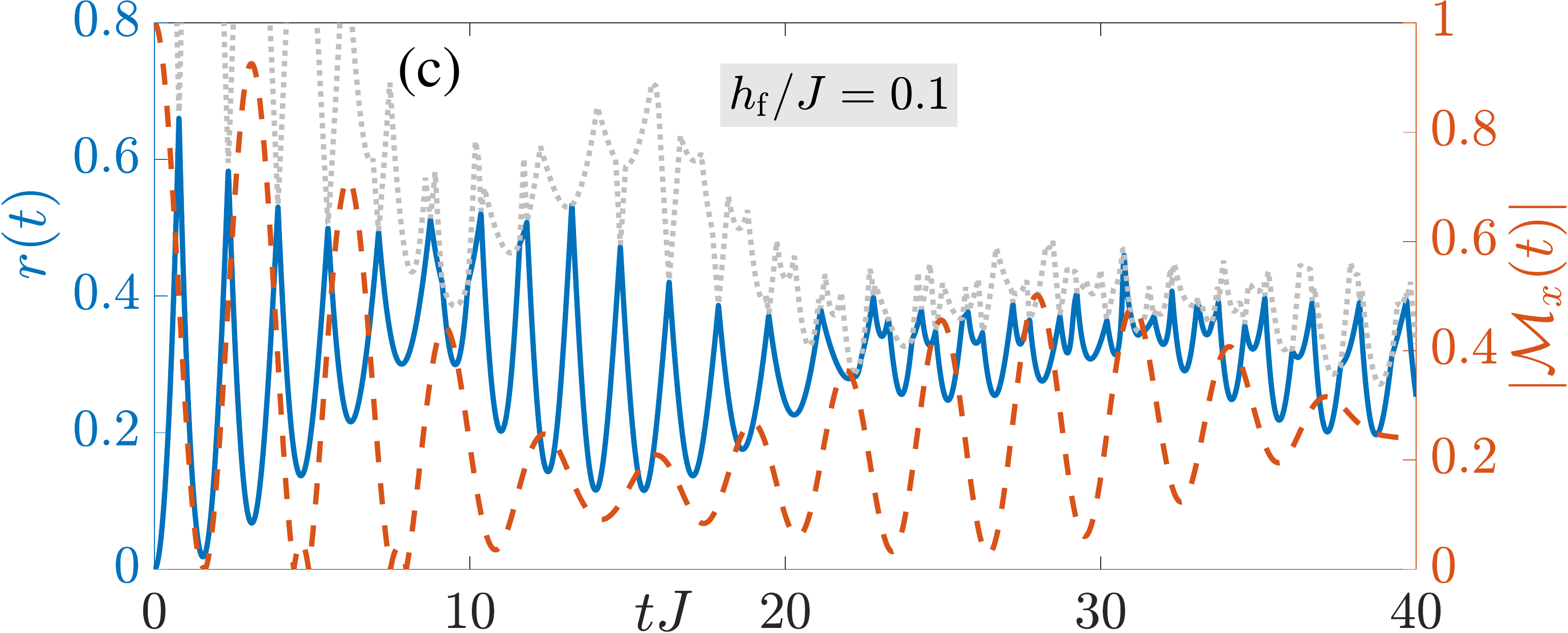}\quad
	\includegraphics[width=.49\textwidth]{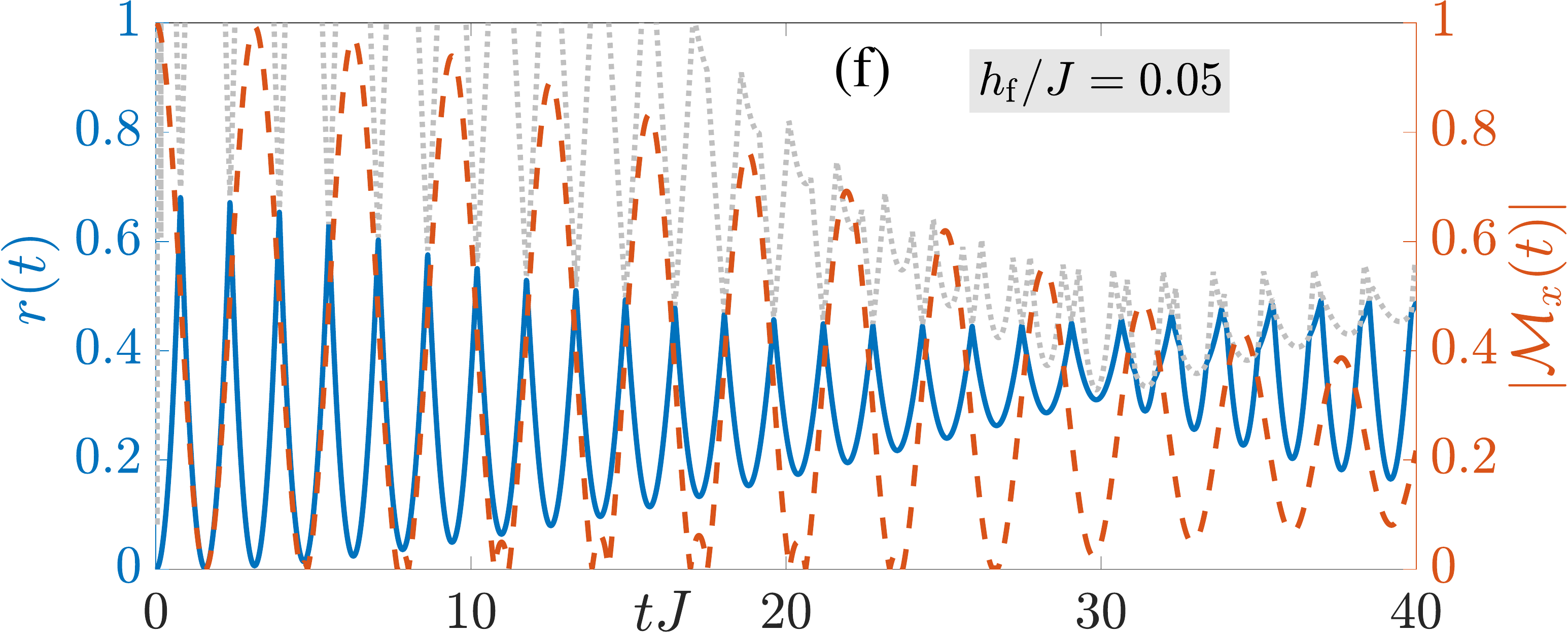}
	\caption{(Color online). Loschmidt return rates after quenching a fully $x$-polarized state with $\hat{H}(h_\mathrm{f})$ for $\lambda=2$ (a-c) and $\lambda=3$ (d-f). Dotted lines indicate sections of the branch cut above the return rate (solid blue line), and cusps form when they intersect. The transverse magnetization is represented by a dashed orange line. Just as in the main text, we see three distinct dynamical phases: (a,d) TDP for $h_\mathrm{f}>h_\mathrm{c}^\mathrm{e}$, (b,c) RDP for $h_\mathrm{f}\in(h_\mathrm{cross},h_\mathrm{c}^\mathrm{e})$, and (c,f) a coexistence region of ADP and RDP for $h_\mathrm{f}<h_\mathrm{cross}$.}
	\label{fig:XQuenchExplamLarge} 
\end{figure*}

\begin{figure}[htp]
	\centering
	\hspace{-.25 cm}
	\includegraphics[width=.49\textwidth]{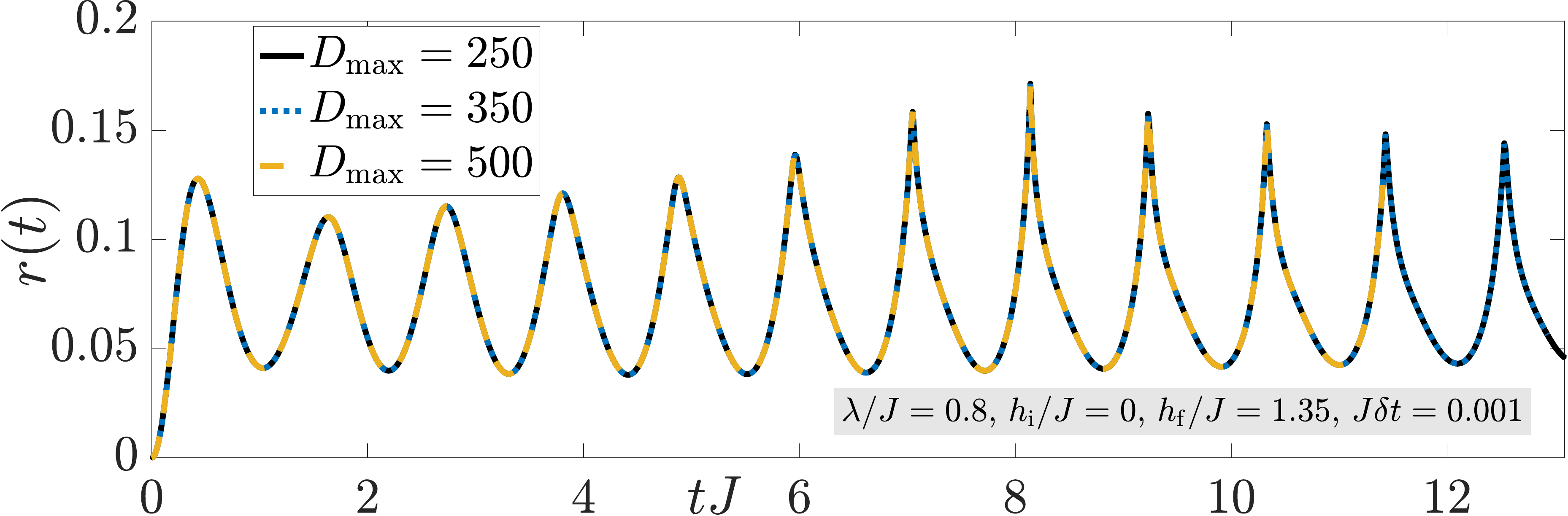}
	\caption{(Color online). Different values of the maximum bond dimension $D_\mathrm{max}$ were used in our simulations, indicating convergence at $D_\mathrm{max}=350$ or lower with a time-step of $\delta t=0.001/J$.}
	\label{fig:convergence} 
\end{figure}

\section{Landau-Lifshitz argument}\label{sec:Landau}
We consider the fully $z$-polarized ground state of the Ising chain described by the Hamiltonian

\begin{align}
H=-J\sum_{j>l=1}^NV(|l-j|)\sigma_l^z\sigma_j^z,
\end{align}
where we take the thermodynamic limit $N\to\infty$ and $J>0$. We now add a ``droplet'' \cite{Landau2013,Thouless1969} of opposite spin polarization along $L$ neighboring sites. The energy cost of this droplet with respect to the ground state is

\begin{align}
\Delta E=2J\sum_{r=1}^N\sum_{l=1}^LV(L+r-l).
\end{align}
Since there are $N$ sites on which this droplet can be positioned, the entropy change is $\Delta S=\ln N$. Thus, the change in the Gibbs free energy upon adding this droplet is

\begin{align}\nonumber
\Delta F&=\Delta E-T\Delta S\\
&=2J\sum_{r=1}^N\sum_{l=1}^LV(L+r-l)-T\ln N.
\end{align}
In the case of nearest-neighbor interactions, $\Delta E=4J$, and the change in free energy becomes $\Delta F=4J-T\ln N$. It is therefore clear that for large $N$, $\Delta F<0$ for $T>0$, which means that the ordered state cannot be the equilibrium state unless $T=0$ for the nearest-neighbor Ising chain.

Let us now consider exponentially decaying interactions $V(r)=\text{e}^{-\lambda (r-1)}$, which means that 

\begin{align}\nonumber
\Delta E&=2J\sum_{r=1}^N\sum_{l=1}^L\text{e}^{-\lambda(L+r-1-l)}\\\nonumber
&=2J\sum_{r=1}^N\text{e}^{-\lambda (r-1)}\sum_{l=1}^L\text{e}^{-\lambda(L-l)}\\
&\overset{N\to\infty}{\to}\frac{2J}{1-\text{e}^{-\lambda}}\sum_{l=0}^{L-1}\text{e}^{-\lambda l}\overset{L\to\infty}{\to}\frac{2J}{(1-\text{e}^{-\lambda})^2},
\end{align}
which shows that in the limit $N\to\infty$, $\Delta F$ will be negative at $T>0$ irrespective of the value of $L$. Hence, an ordered state cannot be the equilibrium state except at $T=0$ for any $\lambda>0$. Therefore, in the case of exponentially decaying interactions there is no finite-temperature phase transition in the Ising chain.

\section{Results for larger $\lambda$}\label{sec:Larger}
In this section we provide additional results supporting the conclusions in the main text. Our initial state is the fully $x$-polarized state, the ground state of $\hat{H}(h_\mathrm{i}\to\infty)$, which we quench with $\hat{H}(h_\mathrm{f})$ of~\eqref{eq:Ham} with $J>0$. Just as in the main results of Figs.~\ref{fig:XQuenchExplam0p4} and~\ref{fig:XQuenchExplam0p8}, in the case of $\lambda=2$ and $\lambda=3$ in Fig.~\ref{fig:XQuenchExplamLarge} we also see three distinct dynamical phases in the return rate~\eqref{eq:RR}. For quenches above the equilibrium critical point $h_\mathrm{c}^\mathrm{e}$, i.e., within the same paramagnetic phase, the return rate is smooth. On the other hand, when $h_\mathrm{f}$ is across $h_\mathrm{c}^\mathrm{e}$ but still above $h_\mathrm{cross}$, the return rate shows regular cusps that are evenly spaced in time. Finally, when $h_\mathrm{f}<h_\mathrm{cross}$, the return rate fundamentally changes with respect to the nearest-neighbor case, where regular cusps appear alongside anomalous cusps that are not evenly spaced in time.

Here it is worth noting that constructing the dynamical phase diagram for $h_\mathrm{i}\to\infty$ is in fact easier than for $h_\mathrm{i}=0$. This is mainly due to the fact that the former makes for much larger quenches, thereby speeding up the dynamics and allowing for anomalous cusps to appear at much earlier times. Indeed, for $h_\mathrm{f}=\delta h\to0^+$, if $h_\mathrm{i}=0$ one would have to wait to $t\to\infty$ in order to witness an anomalous cusp in the return rate, because this is an infinitesimally small quench. On the other hand, if $h_\mathrm{i}\to\infty$, then this is a huge quench, which inevitably leads to the appearance of cusps at much earlier times in the return rate.

\section{Convergence}\label{sec:Convergence}
For our numerical simulations, we find that all results converge at maximum bond dimension $D_\mathrm{max}=350$ and time-step $\delta t=0.001/J$. In Fig.~\ref{fig:convergence}, we show the return rate for a quench on the fully $z$-polarized state with $h_\mathrm{f}=1.35J$ and $\lambda=0.8$, which displays ADP for different values of $D_\mathrm{max}$ indicating convergence.

\end{document}